# Gravitational effect on the advancing and receding angle of a 2D Cassie-Baxter droplet on a textured surface

*Donggyu Kim[1], Keonwook Kang[2], Seunghwa Ryu[1,\*]*

[1]Department of Mechanical Engineering, Korea Advanced Institute of Science and Technology (KAIST), Daejeon 34141, Republic of Korea

[2]Department of Mechanical Engineering, Yonsei University, Seoul 03722, Republic of Korea

Corresponding author Fax: +82-42-350-3059

E-mail address: ryush@kaist.ac.kr

**Abstract**

Advancing and receding angles are physical quantities frequently measured to characterize the wetting properties of a rough surface. Thermodynamically, the advancing and receding angles are often interpreted as the maximum and minimum contact angles that can be formed by a droplet without losing its stability. Despite intensive research on wetting of rough surfaces, the gravitational effect on these angles has been overlooked because most studies have considered droplets smaller than the capillary length. In this study, however, by combining theoretical and numerical modeling, we show that the shape of a droplet smaller than the capillary length can be substantially modified by gravity under advancing and receding conditions. First, based on the Laplace pressure equation, we predict the shape of a two-dimensional Cassie-Baxter droplet on a textured surface with gravity at each pinning point. Then the stability of the droplet is tested by examining the interference between the liquid surface and neighboring pillars as well as analyzing the free energy change upon depinning. Interestingly, it turns out that the apparent contact angles under advancing and receding conditions are not affected by gravity, while the overall shape of a droplet and the position of the pinning point is affected by gravity. In addition, the advancing and receding of the droplet with continuously increasing or decreasing volume is analyzed, and it is showed that the gravitational effect plays a key role on the movement of the droplet tip. Finally, the theoretical predictions were validated against line-tension based front tracking modeling that seamlessly captures the attachment and detachment between the liquid surface and the solid substrate. Our findings provide a deeper understanding on the advancing and receding phenomena of a droplet, and essential insight on designing devices that utilize the wettability of rough surfaces.

# 1. Introduction

The advancing and receding angles are important physical quantities that have been extensively used to characterize the wettability of textured surfaces (1-5). Because a droplet can have multiple local minima when it sits on a chemically or topographically heterogeneous surface (6-8), one would measure many different contact angles that are determined by its initial deposition location, deposition speed, and ambient vibration (9,10). Hence, instead of a single contact angle, researchers in the field use two contact angles (the contact angles when the triple line of the droplet starts to advance and recede) as a characteristic feature of the wettability of the surface (2,3,11-14). For example, contact angle hysteresis (CAH) (15-17), defined as the difference between the advancing and receding angles, has been used as a measure of the liquid-solid cohesion on a rough surface: a droplet sticks well to a surface with high CAH, while it slips well on a surface with a small CAH (18-20). Thermodynamically, the advancing and the receding angles have been interpreted as the maximum and minimum contact angles that can be formed by a droplet without losing its stability (2,21,22). An analysis on the free energy landscape of the Cassie-Baxter (CB) droplet (22-25) on a textured surface revealed that a droplet can maintain its shape as long as the contact angle is larger than the Young's angle of the substrate ($\theta_e$), i.e., the contact angle of a perfectly flat substrate (9).

Existing theoretical analyses led to the conclusion that an arbitrarily large CB droplet with a contact angle larger than the Young's angle can sustain metastable pinning. However, such an argument is contradictory because the advancing angle has predominantly been measured by recording the advancement of a droplet upon its increase in volume (1,5). The discrepancy has been attributed to external effect such as ambient vibration that makes the droplet overcome a relatively small energy barrier to de-pin from an initial pining point (10,26,27). However, the primary source of the external perturbation as well as its magnitude has neither been thoroughly analyzed nor suggested with verification. Therefore, we could raise a few questions: Is the ambient vibration the primary source causing the instability of a droplet? Can a droplet maintain its pinning point when the volume of the droplet is significantly increased if the measurement is performed by completely isolating external perturbation such as vibration?

In this study, we rigorously analyze the gravitational effect on a droplet shape under advancing and receding conditions and show that gravity can have a significant effect on stability, even for a droplet much smaller than the capillary length ($\lambda_C$). First, the shape of a droplet under gravity at each of the pinning points is predicted by solving the Laplace pressure equation using a numerical method (28) (**Fig. 1a, b**). Afterward, the stability of a droplet is determined by examining the interference between the deformed droplet and neighboring pillars, and the free energy change upon depinning (**Fig. 1c**). Based on these analyses, we confirmed that the pinning points under advancing and receding conditions and the droplet shape can be substantially affected by the gravitational effect, and the gravitational effect can explain the advancement of the droplet tip of the Cassie-Baxter state without help of the external effect. Finally, theoretical predictions are validated against the results from the line-tension based front tracking modeling (LTM) that seamlessly captures the attachment and detachment phenomenon between the liquid interface and the substrate surface.

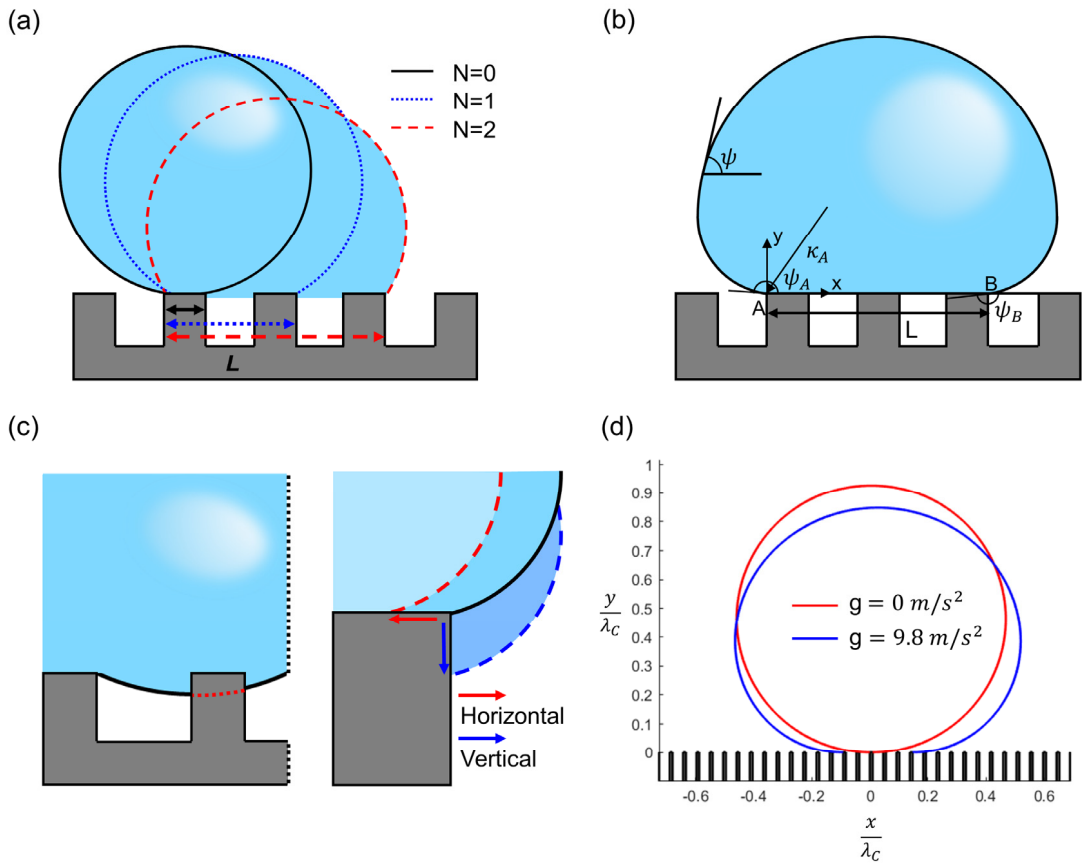

**Figure 1. Gravitational effect on advancing and receding droplets**. (a) Pinning point of a Cassie-Baxter droplet on a textured surface (b) Modeling of a droplet deformed by gravity (c) Method to check if the droplet can maintain the shape at the pinning point. Interference check (left) and free energy analysis at depinning (right) (d) Droplet shape under advancing condition without (red solid) and with gravity (blue solid).

## 2. Methods

### 2.1. Geometrical stability of the droplet

We first obtain the droplet shapes at every pinning point of the Cassie-Baxter droplet on a textured hydrophobic surface in the absence of gravity. It is known that the tips of such a CB droplet are located at the outer ends of the pillars (**Fig.1a**) (9). The basal length (L) of a droplet under each pinning condition is related to other geometrical factors such as $L = N(W + G) + W$, where W, G, and N refer to the width of the step, the groove of the textured geometry, and the number of grooves below the droplet, respectively. We choose N to define a pinning state and use the angle formed by the undeformed droplet (i.e., the droplet shape that does not incorporate the deformation by gravity), which is subsequently referred to as a zero-g angle ($\theta^0$ in **Fig. 2a**), to define the contact angle at the pinning point, which it turns out, best represents the overall shape of the droplet. A small zero-g angle indicates a wider basal length and *vice versa*. As we will show later, the apparent contact angles of a deformed droplet ($\theta^1$ in **Fig. 2a**) under advancing and receding conditions are not affected by gravity, and hence $\theta^1$ cannot serve as a metric representing the overall shape. At each pinning point (N = 0,1,2,…), the shape of the droplet under gravity is predicted by numerically solving the Laplace pressure equation (28). The contour of the droplet can be determined with the followed differential equation. (28)

$$\frac{dX}{d\psi} = \frac{\cos\psi}{Q}, \frac{dY}{d\psi} = \frac{\sin\psi}{Q}, \text{where } Q = \frac{\rho g}{\gamma_{LV}} y - \kappa_A$$

where $\psi$ refers to the angle between the horizontal plane and the position of the liquid surface; X and Y refer to the horizontal and vertical coordinates of the droplet contour, respectively; and $\kappa_A$ refers to the local curvature at point A (**Fig. 1b**). The differential equation can be solved using the 4$^{th}$ order Runge-Kutta method by assigning the appropriate initial conditions for the angles at each end of the droplet: $\psi_A$, $\psi_B$, and $\kappa_A$. The Newton Raphson method was employed to find the initial conditions ($\psi_A, \psi_B, \kappa_A$) that satisfy the given droplet volume and the pinning position (See supplementary Information).

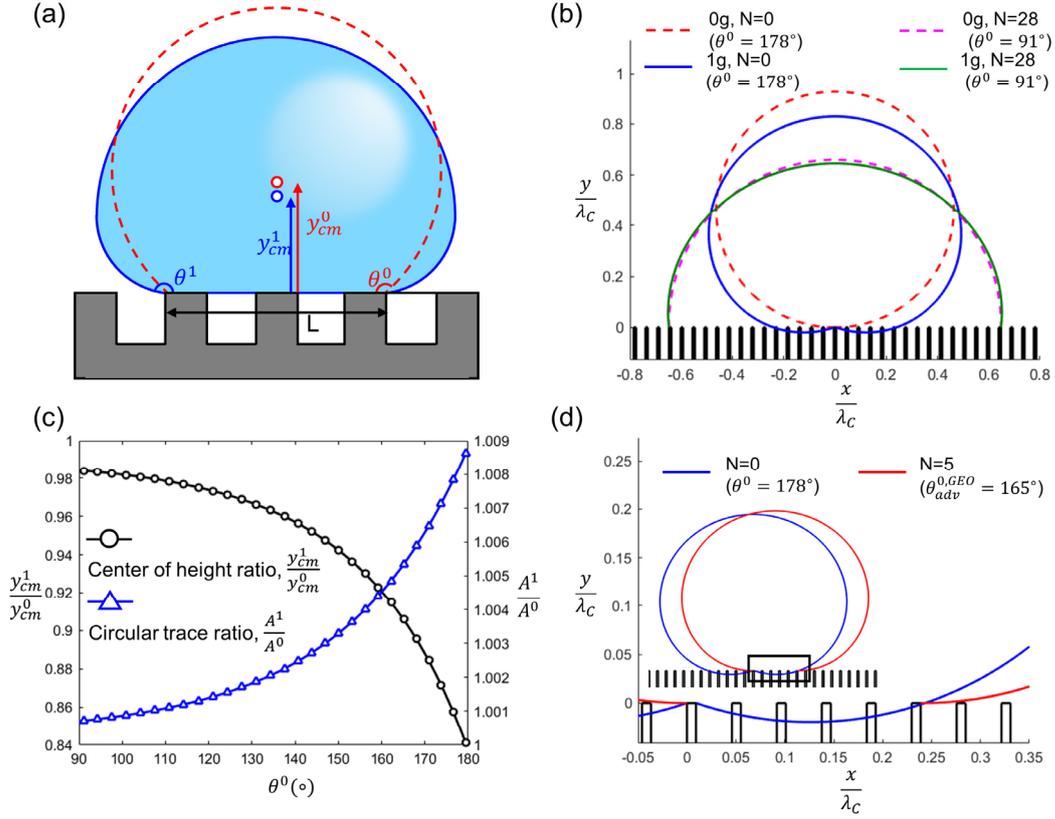

**Figure 2. Gravitational shape change of the droplet on a rough surface.** (a) Model of the droplet before (red, dotted) and after (blue, solid) gravitational effect is considered. (b) The shape of the droplet when the droplet has a large zero-g angle ($\theta^0 = 178°$, blue and red), and moderate value ($\theta^0 = 91°$, green, purple). The droplet shows larger deformation with a large zero-g angle. (c) Change of height of the center of mass and surface area of the circular trace due to gravity (d) The deformed droplet possessing a different number of grooves below. When the droplet has a small number of grooves below, the interference between the liquid surface and pillars can be formed.

We then analyze how the droplet shape at each pinning point is deformed by gravity. Fig. 2b shows the configuration of the water droplet with a volume of $V = 5 \times 10^6 \mu m^2$ before (red, purple dotted) and after the shape change due to gravity (blue, green solid). Here the radius of the circular droplet before being placed on the surface is about half the capillary length ($R_0 = \sqrt{\frac{V}{\pi}} = 0.47\lambda_C$) and the Young's angle is $\theta_e = 95°$ on the flat surface. As depicted in **Fig. 2b**, the gravity-induced deformation is larger when it has an extremely small basal length (L = W, $\theta^0 = 178°$), and it becomes smaller when the droplet has a larger basal length (L = 29W + 28G, $\theta^0 = 91°$). To illustrate the deformation of the droplet, relative changes in the center of mass location $\left(\frac{y_{cm}^1}{y_{cm}^0}\right)$ and the surface area of the droplet cap $\left(\frac{A^1}{A^0}\right)$ are plotted in terms of the zero-g angle, as shown in **Fig 2c**. In all cases, the difference increases for a droplet having a higher zero-g contact angle (i.e., smaller basal length). Within the limit of an almost 180° zero-g angle, the droplet shows a 15% difference in its center-of-mass height, while the boundary area of the cap (i.e., arc length for a 2D droplet) varies by only 1%. Based on the results, we can conclude that the shape of a droplet smaller than the capillary length changes significantly due to gravitational effect while the surface area is kept nearly unchanged.

The contour of a droplet deformed by gravity may interfere with the neighboring pillars (blue solid in **Fig. 2d**). Such interference cannot be seen when gravity is neglected because the droplet contour must be part of a circle. When a droplet makes contact with a neighboring pillar, it cannot maintain its shape at the pinning point. Hence, we obtain the minimum basal length L (minimum number of N) above which such interference disappears. For instance, the large zero-g contact angle droplet (blue solid in **Fig. 2d**) with N = 0 interferes with pillars, while an identically sized droplet does not make interference when N = 5 (red solid in **Fig. 2d**). Through this process, we can predict the geometrically determined stable droplet shape under advancing condition, when the volume of the droplet and the substrate geometry is given.

## 2.2. Free energy based stability of the droplet

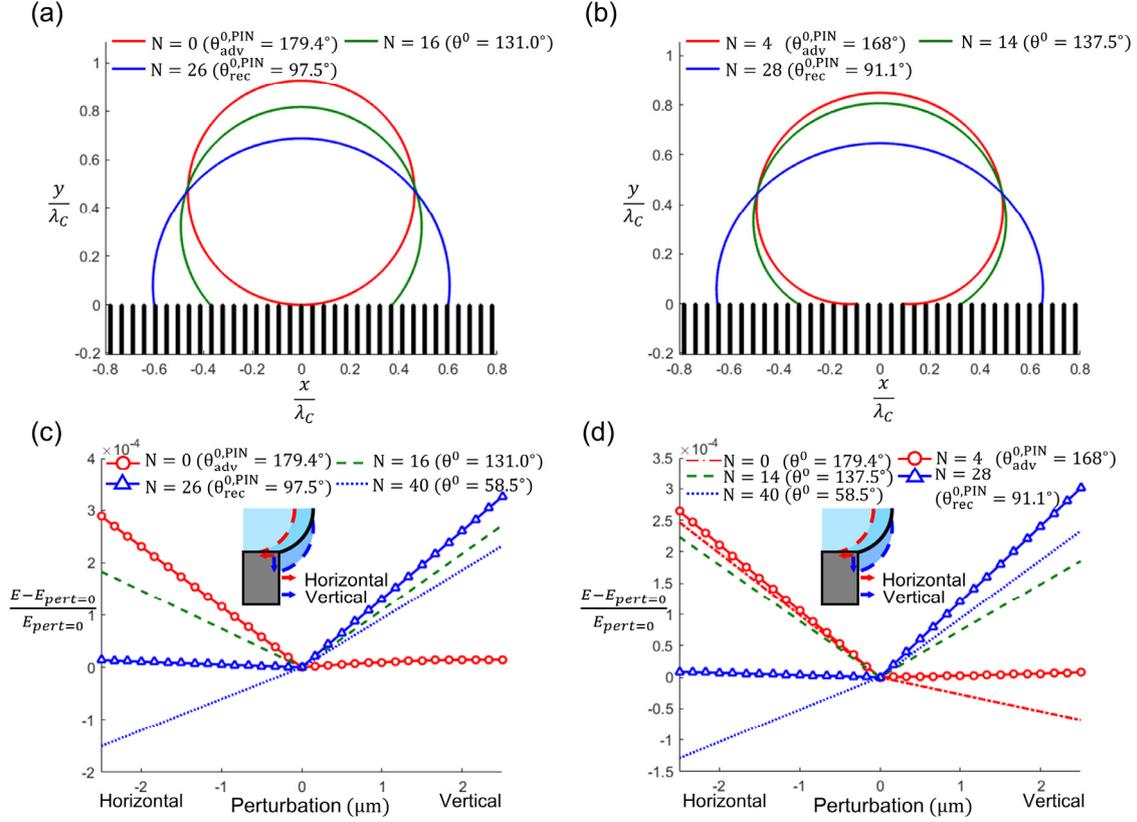

**Figure 3. Pinning stability of the droplet.** (a, c) The shape of the droplet at each of the pinning points with 0g and the free energy curve upon depinning at the corresponding pinning point. (b, d) The shape of the droplet at each pinning point with 1g and the free energy curve upon depinning at the corresponding pinning point. Red and blue solid with a symbol each refers to the advancing and receding angle from the pinning stability condition.

Avoiding interferences does not automatically guarantee the stability of a droplet because the tip of a droplet can slide if the free energy is lowered upon the depinning. We compute the free energy change upon small perturbations on the pining point to determine the stability of a droplet, while assuming the flat bottom surface of the Cassie-Baxter state. The free energy of a droplet can be expressed as:

$$E = \gamma_{LV}(A_{LV} - A_{SL}\cos\theta_e) + \rho V g y_{cm}$$

where $A_{LV}$ and $A_{LS}$ refer to the surface area between liquid and vapor, and liquid and substrate, respectively. We fix one end of a droplet at the pining point and compute the free energy change upon horizontal and vertical perturbations of the other end (inset of **Fig. 3c,d**) in the presence and absence of gravity. In this example, we consider a water droplet with a $5 \times 10^6 \mu m^2$ volume ($R_0 = \sqrt{\frac{V}{\pi}} = 0.47\lambda_C$) on a substrate with a 25 μm width (W), 100 μm groove length (G), and a Young's angle of $\theta_e = 95°$. When the free energy curve has its local energy minimum at zero perturbation, the tip of the droplet will stay at the pinning point. However, if the energy curve decreases upon either horizontal or vertical perturbation, the tip would slide from the pinning point to reach the nearest energy minimum. In the absence of gravity, we find that even a droplet sitting on only one pillar (N = 0, red solid with symbol) is stable upon depinning (**Fig. 3c**) despite its very large zero-g contact angle of about 180°. With an increase in the basal length, the slope of one side of the energy curve decreases (green dotted), and eventually reaches negative (i.e., the droplet loses its stability) when the zero-g contact angle becomes smaller than the Young's angle of the substrate (blue dotted). Therefore, in the absence of gravity, we can conclude that the advancing zero-g angle would become nearly 180°, and the receding zero-g angle is around the Young's angle of the substrate (blue solid curve with symbol), which is consistent with existing research (9). On the other hand, when gravitational effect is considered, the droplet does not show stable free energy behavior for very small basal lengths (N = 0, red dot dashed in **Fig. 3d**). Here, the maximum stable zero-g contact angle turns out to be $\theta^0 = 168°$ at N = 4, which can also called as the pinning based advancing angle in zero-g contact angle ($\theta_{adv}^{0,PIN}$). A droplet sustains its stability until the zero-g contact angle reaches the value smaller than the Young's angle ( N = 28, blue solid with symbol, $\theta_{REC}^{0,PIN} = 91.1°$). Based on the results, we discover that gravity lowers both the advancing and receding angles (defined in terms of the zero-g angle), forcing a droplet to sit on more pillars, even if the droplet is smaller than the capillary length.

## 2.3 Line-tension based front tracking modelling (LTM)

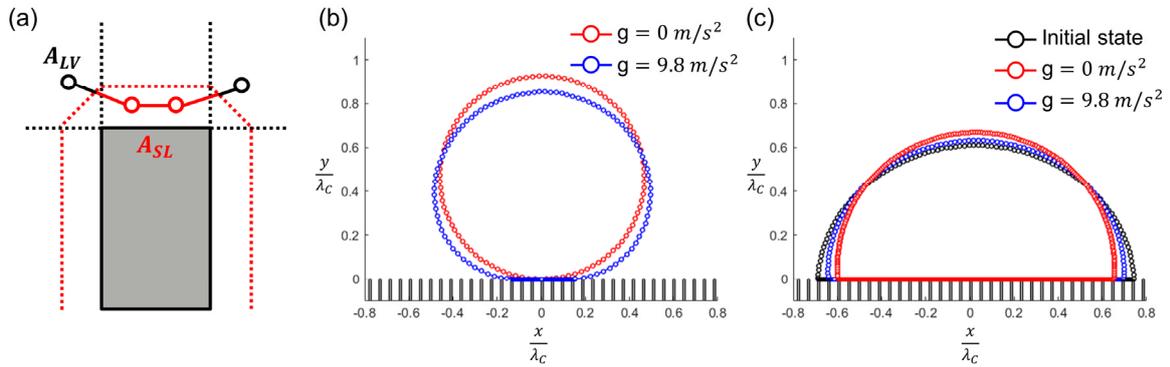

**Figure 4. Line tension-based front tracking modeling (LTM) for predicting the advancing and receding angle of the droplet** (a) Element modeling for the LTM. The element near the textured geometry is modeled as the surface between liquid and substrate (red), and the others are modeled as the surface between liquid and vapor (black). (b) Droplet configuration under an advancing condition, with and without gravity. (c) Droplet configuration under a receding condition, with and without gravity

Next, the theoretically predicted advancing and receding angles of a droplet on a textured surface are numerically validated against the results from the line-tension based front tracking modeling (LTM) which is a numerical simulation that can seamlessly model the attachment and detachment between the liquid interface and the substrate surface. In the present study, we chose to validate the theoretical predictions against the numerical simulation results instead of the experimental measurement because of two reasons. First, it was recently reported that it is very hard to precisely locate the pinning point of the droplet with a typical goniometer because the distance between the flatted droplet contour and the pillars next to the pinning point is very small and thus small errors in contact angle measurements can lead to a significant error in determining pinning point position (29). Second, although our theoretical analysis is based on two-dimensional droplet for the sake of mathematical simplicity (it is infeasible to derive exact analytical expressions for three-dimensional droplets), the experimental observation should be conducted with three-dimensional droplet. Considering the complexity of making quasi-2d droplet experiments, combined with the limitation of experimental resolution mentioned above, we concluded that it is very hard to draw a conclusion from the direct comparison between the theoretical prediction and experimental measurements.

Because of aforementioned reasons, we chose to validate the theoretical results with the line-tension based front tracking method, a virtual experimental tool developed for the present work. In the frame work, the liquid interface is modeled as a chain of nodes connected by line segment elements; the elements near the substrate represent the liquid-substrate interface, and the others represent the liquid-vapor interface. To implement it, the surface tension function which gradually varies according to the distance from the substrate geometry was introduced (**Fig. 4a**). The droplet shape is then determined by minimizing the free energy while satisfying the constraints for the augmented Lagrangian method (30). We emphasize that the adaptation of the gradually varying surface tension function is the key enabler for modeling the attachment and detachment, which is infeasible in most existing numerical methods adapting a Hessian function type surface tension. A detailed description of the method is provided in the Supplementary Information. Because the numerical simulation does not pose any assumption on the bottom surface shape of the Cassie-Baxter state (as is the case for the theoretical model), the simulation model captures the sagging of the droplet contour between the pillars (31). Also, while the theoretical analysis assumes that the contour of the droplet follows the Laplace equation, the simulation model evolves by a simple principle of energy minimization subjected to constant liquid volume constraint without such assumption and thus provides an independent prediction on the droplet shape.

To predict the advancing and receding angles of a droplet, we consider droplets on a pining point with different contact angles as initial configurations. To obtain the advancing angle, we choose the droplet configuration sitting on only one pillar below (N = 0) as the initial geometry (red in **Fig. 4b**) and relax the droplet shape via energy minimization. Consistent with the theory, the droplet shape is maintained at the initial shape in the absence of gravity. In contrast, with a gradual increase in the gravitational effect, the shape of the droplet changes, with the basal length increasing (i.e., a higher number of grooves below) until it reaches the droplet configuration under advancing condition predicted from the theoretical analyses (blue in **Fig. 4b**). On the other hand, to obtain a receding droplet, the droplet at the pinning point possessing a contact angle near 80° was prepared as the initial configuration (black in **Fig. 4c**). In the absence of gravity, the tip of the droplet slides until it forms the contact angle near the Young's angle upon the energy minimization step. With the gravitation effect, the droplet settles at a configuration sitting on more pillars than in the case of zero-gravity. Our numerical modeling confirms that the stable droplet shape under gravity has a larger basal length under both advancing and receding conditions than the basal length predicted without gravity.

# 3. Results and Discussion

## 3.1 Volumetric effect on the advancing and receding contact angle

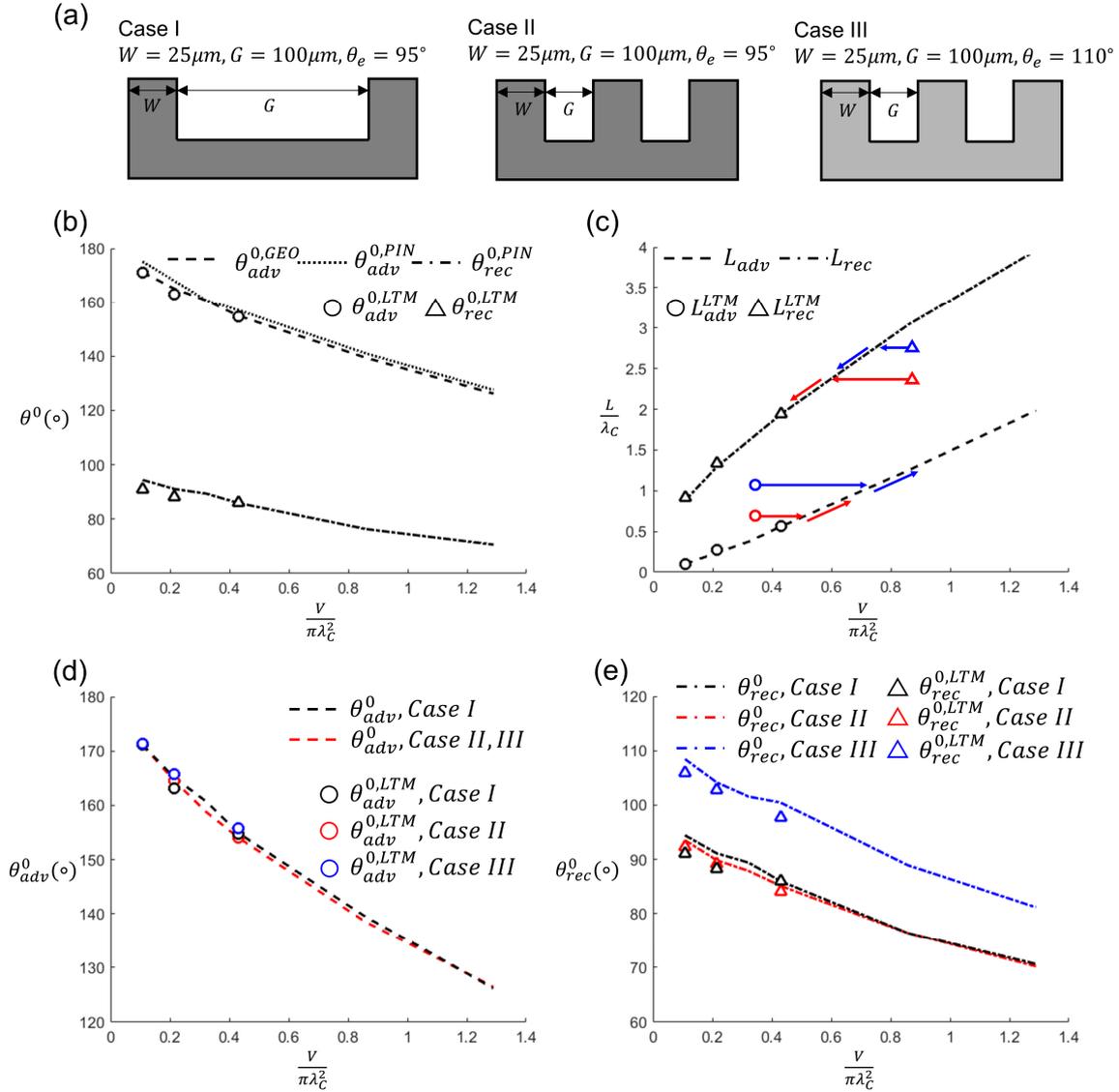

**Figure 5. Effect of Young's angle and geometry of the substrate on the advancing and receding angles.** (a) Three cases with different substrate conditions were considered (Case I: ($\boldsymbol{\theta_e = 95°, W = 25\mu m, G = 100\mu m}$), Case II: ($\boldsymbol{\theta_e = 95°, W = 25\mu m, G = 25\mu m}$), and Case III: ($\boldsymbol{\theta_e = 110°, W = 25\mu m, G = 25\mu m}$)). (b) Predicted advancing and receding zero-g angle for Case I: $\boldsymbol{W = 25\mu m, G = 100\mu m, \theta_e = 95°}$. (c) Predicted basal length of the droplet at advancing and receding condition (d) predicted advancing zero-g angle at each of the substrate cases, and (e) predicted receding zero-g angle at each of the substrate cases

First, the volumetric effect on the advancing and receding contact angle is presented with the substrate condition, $\theta_e = 95°, W = 25\mu m, G = 100\mu m$. For the analysis, a wide range of volume, from $V = 2.5 \times 10^6$ ($R_0 = 0.33\lambda_C$) to $3 \times 10^7 \mu m^2 (R_0 = 1.13\lambda_C)$ is considered. **Fig. 5b** presents the predicted zero-g angles under advancing and receding conditions on the substrate condition. The dotted line refers to the geometrically determined advancing angle, the dashed line refers to the pinning stability based advancing angle, the dot-dashed line refers to the pinning based receding angle, and the symbols refer to the advancing and receding angles from the LTM. Because the droplet should satisfy both the geometrical and pinning stability conditions to maintain the shape, the smaller advancing angle of the geometrically determined $\theta_{adv}^{0,GEO}$ and pinning stability based $\theta_{adv}^{0,PIN}$ is chosen as the zero-g angle under the advancing condition, while the receding angle is determined solely through the pinning stability criterion as $\theta_{rec}^{0,PIN}$. The results shows that the theoretical predictions match well with the results from the LTM under both advancing and receding conditions. Interestingly, the advancing angle decreases from $180°$ and the receding angle decreases from the Young's angle of the surface with an increase in volume, and the rate of change is substantial even in the small volume range (a droplet significantly smaller than the capillary length).

Also, we analyze the advancing and receding of the droplet tip upon gradual volume change. **Fig. 5c** shows the how the normalized basal length of the droplet (with respect to the capillary length) at advancing and receding condition varies upon the normalized volume change. The dashed line refers to the normalized basal length at advancing condition, the dot-dashed line refers the basal length at receding condition, and black circle and triangle symbols refer to the LTM results. It is shown that the basal length at both advancing and receding condition increases as the droplet volume increases, and the droplet having intermediate basal length between two limiting lengths at advancing and receding conditions can be considered as a metastable one. If the volume of the droplet which was initially at a meta-stable state (for instance, blue circle in **Fig. 5c**) increases gradually, it maintains its original pinning point, preserving the basal length in the early stage. With more volume increase, the basal length meets the advancing condition (balck dashed). Since the droplet cannot find the stable state below the advancing state boundary, the droplet tip advances and the droplet follows the dashed line upon further volume increase, which means the continuous advancing of the droplet tip. The receding droplet with gradual volume reduction (red triangle) would show a similar behavior. The theoretically predicted droplet behavior is consistent with the typical droplet tip advancing or receding experiment which shows constant basal length at the early stage and the continuous advancing or receding phenomena after the tip starts to move (32,33). These behaviors cannot be explained with the existing theoretical frameworks that do not account for gravitational effect because there is no lower limit of basal length for the stable droplet, and thus external stimuli such as ambient vibration has been considered as the main cause of the advancement in the existing studies. The theoretical model in the present study shows the gravitational effect can trigger the advancement of the droplet tip without any aid of external stimuli such as ambient vibration. Another interesting prediction is that two droplets with identical volume but different initial pinning point (blue and red circles, blue and red triangles in **Fig. 5c**) would start to advance or recede at different volume. It shows the droplet configuration at advancing and receding condition is not only affected by the volume, but also is affected by the initial pinning point of the droplet.

## 3.2 Substrate chemistry and the texture geometry effect on the advancing and receding contact angle

Finally, we analyze in detail the effects of the Young's angle (i.e., different substrate material) and the substrate geometry on the advancing and receding angles. The three different substrate geometries are considered (**Fig. 5a**); Case I: ($\theta_e = 95°, W = 25\mu m, G = 100\mu m$), Case II: ($\theta_e = 95°, W = 25\mu m, G = 25\mu m$), and Case III: ($\theta_e = 110°, W = 25\mu m, G = $

$25\mu m$). The effect of surface texture would manifest in the difference between Cases I and II, while the effect of surface chemistry would manifest in the difference between Cases II and III. **Fig.5 d,e** presents how the advancing and receding angles is affected by the substrate conditions. We find that the geometrically decided advancing angle is chosen as the advancing angle for all cases considered here and the advancing angles for all three cases turn out to be very similar to each other. Specifically, Case II and Case III, which have identical surface textures, show precisely the same theoretically predicted advancing angle. This is somewhat expected because the geometrically determined advancing angles $\left(\theta_{adv}^{0,GEO}\right)$ of all three cases are smaller than the pinning stability determined advancing angles $\left(\theta_{adv}^{0,PIN}\right)$, i.e., the advancing angles are determined by geometrical interference between the droplet contour and neighboring pillars. In essence, the result indicates that the advancing angle is primarily determined by the surface texture rather than the surface chemistry. Yet, the effect of surface texture is rather limited, and the difference between the advancing angles of Cases I and II is small. On the other hand, the receding angle turns out to be determined primarily by the surface chemistry as indicated by the almost identical receding angles of Cases I and II, and the very different values in Cases II and III. This is because the receding angle is determined by the free energy change upon small depinning perturbation that depends on the surface chemistry (i.e., solid-vapor and solid-liquid interfacial energies). Overall, it is interesting to note that the effect of surface texture on both advancing and receding angles is rather limited. In all cases, while the zero-g angles ($\theta^0$ in **Fig. 2a**) under advancing and receding conditions are affected by gravity, the apparent angles ($\theta^1$ in **Fig. 2a**) are not affected by gravity and remain the same at $180°$ and $\theta_e$ for the entire range of droplet volume. A detailed discussion on the apparent advancing and receding angels is presented in the Supplementary Information.

## 4. Conclusion

There have been a plethora of research to tune wetting properties with surface microstructure to achieve special functionality such as self cleaning, anti-fogging, anti-icing, and droplet transport (34-41). Since it is known the advancing and receding angle of the droplet is deeply related with the wetting phenomena, theses angles as well as contact angle hysteresis have been widely used to characterize the wettability of a microstructured surface.

Theoretically, the advancing and receding angles has been considered as the maximum and minimum contact angle that a droplet can possess on the surface (2,10,21,22). For Cassie-Baxter state, all pinning points with contact angle larger than the Young's angle have been known to be metastable state based on the free energy analysis that does not account for the gravitational effect (9). Consequently, the advancement of the droplet tip with finite volume could not be explained with the existing theory, and external effect such as an ambient vibration has been suggested as the origin of the droplet advancement (10,26,27).

In this research, we investigated how the gravity affects the advancing and receding angles of a CB droplet on textured surfaces by appropriately incorporating the gravity. Especially, it has been long believed that gravitational effect can be neglected for droplets smaller than the capillary length (7,12,21,22,42). Nonetheless, by examining geometrical interference and pinning stability, we revealed that the gravitational effect on the droplet configuration under advancing and receding conditions is substantial even for a droplet smaller than the capillary length, and the effect can explain the advancement of the droplet tip without introducing the external effect such as ambient vibration. Finally, the theoretical results were validated with an independent virtual experiment, a numerical simulation based on line tension model. Our study provides a fundamental understanding on the advancing and receding phenomenon and enables precise prediction on droplet configuration under the advancing and receding conditions. We believe the outcome of this study can be further utilized to predict the wettability of a microstructured surface and to design the surface with required wettability.

# Gravitational effect on the advancing and receding angle of 2D Cassie-Baxter droplet on textured surface

*Donggyu Kim[1], Keonwook Kang[2], Seunghwa Ryu[1,\*]*

[1]Department of Mechanical Engineering, Korea Advanced Institute of Science and Technology (KAIST), Daejeon 34141, Republic of Korea

[2]Department of Mechanical Engineering, Yonsei University, Seoul 03722, Republic of Korea

Corresponding author. Fax: +82-42-350-3059

E-mail address: ryush@kaist.ac.kr

**Supplementary Note 1: Method to predict the shape of 2D Cassie-Baxter Droplet at pinning point.**

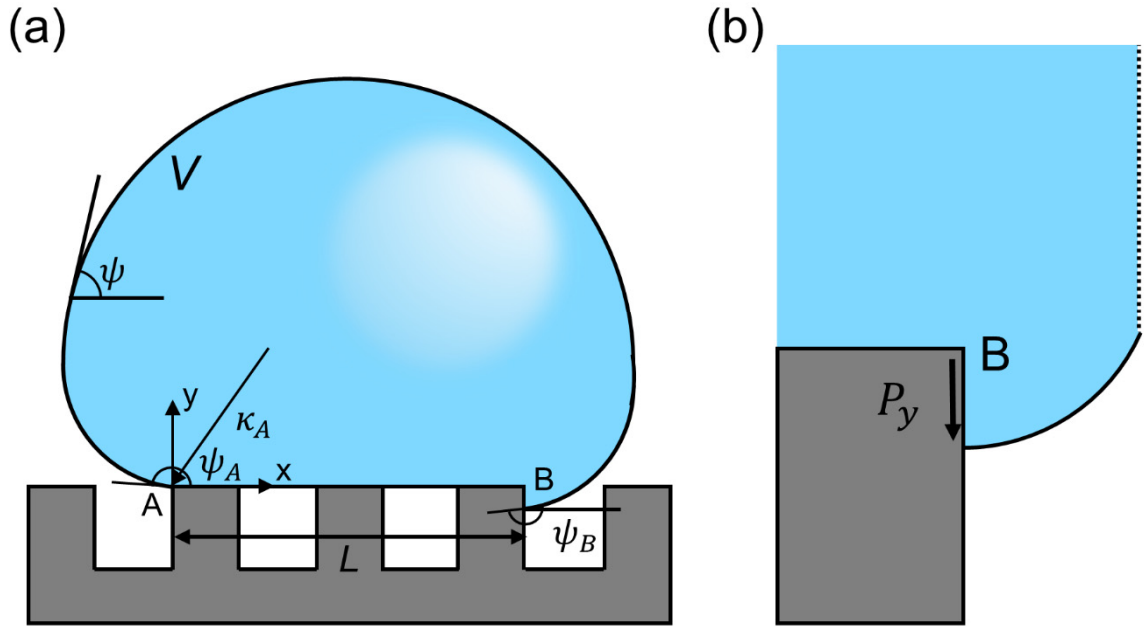

**Figure S1. Modelling of droplet with gravitational effect.** (a) Variables in the modelling. (b) Droplet contour near the tip

We can predict the shape of the droplet with the gravity when the volume of the droplet and the pinning point is given, using the following differential equation[1].

$$\frac{dX}{d\psi} = \frac{\cos\psi}{Q}, \frac{dY}{d\psi} = \frac{\sin\psi}{Q}, \text{where } Q = \frac{\rho g}{\gamma_{LV}} y - \kappa_A$$

where $\psi$ refers to the angle between horizontal plane and the position of the liquid surface. The differential equation can be solved via the 4th order Runge-kuta method using appropriate initial conditions of $\psi_A, \psi_B, \kappa_A$. $\psi_A$ and $\psi_B$ are the contact angles of two ends of the droplet, and $\kappa_A$ is the local curvature of the droplet at the tip A.

The Newton-Raphson method is then employed to find the initial condition **S** = $(\psi_A, \psi_B, \kappa_A)$, which satisfies the given droplet condition, i.e. the volume and the pinning position. First, the initial condition satisfying the given droplet condition at zero gravity was calculated from the geometrical relation. After, by slowly augmenting the gravity, the error function is calculated.

$$f(S) = [V(S) - V_0, L(S) - L_0, P_y(S) - P_{y0}]^T$$

Here, the $V(S)$ refers the volume of the droplet with the initial condition $S$, and $L(S), P_y(S)$ refer the basal length and the height difference between the pinning point (**Fig. S1b**), respectively. They can be calculated as follow.

$$V = \int Y dX$$

$$L = X(\psi_B) - X(\psi_A)$$

$$P_y = Y(\psi_B) - Y(\psi_A).$$

On the other hand, the $V_0, L_0, P_{y0}$ are the values of the given droplet condition. From the error function set, the jacobian matrix $J$ can be numerically formulated,

$$J(S) = \begin{bmatrix} \frac{\partial f_1(S)}{\partial S_1} & \frac{\partial f_1(S)}{\partial S_2} & \frac{\partial f_1(S)}{\partial S_3} \\ \frac{\partial f_2(S)}{\partial S_1} & \frac{\partial f_2(S)}{\partial S_2} & \frac{\partial f_2(S)}{\partial S_3} \\ \frac{\partial f_3(S)}{\partial S_1} & \frac{\partial f_3(S)}{\partial S_2} & \frac{\partial f_3(S)}{\partial S_3} \end{bmatrix}$$

where $S_1, S_2, S_3, f_1, f_2, f_3$ refers to the element of $S$ and $f(S)$ each. Then the initial condition set **S** can be updated as below, until the norm of the error function decays to the value small enough.

$$S_{i+1} = S_i - J(S_i)^{-1} f(S_i)$$

Through the process, the gravitational acceleration is sequentially increased to guarantee the convergence.

# Supplementary Note 2: Line tension based front-tracking method for 2D CB droplet on textured surface

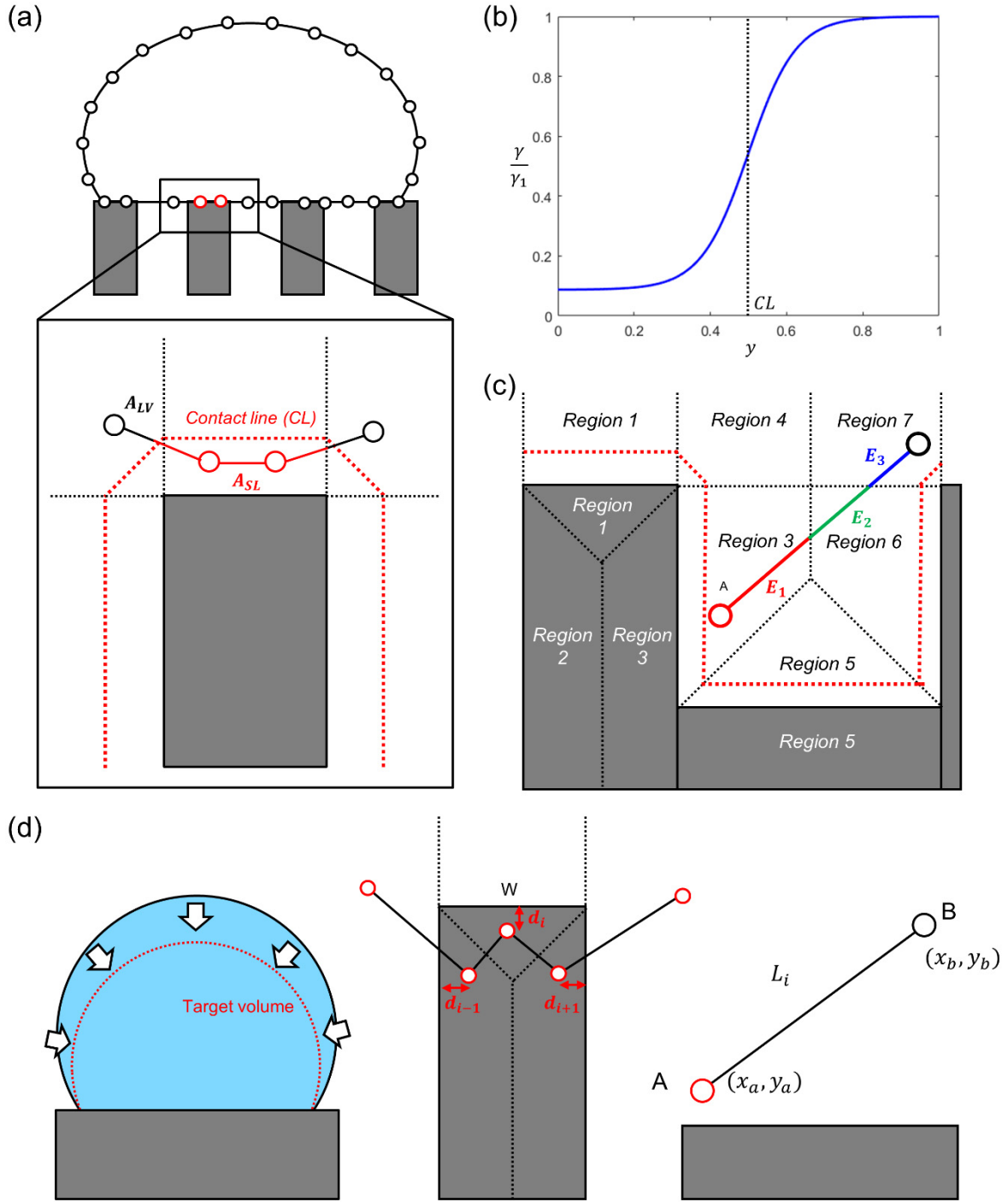

**Figure S2. Line tension based front-tracking method**. (a) Modelling of droplet element. The elements near the substrate refer the liquid-substrate interface, while the others refer the liquid-vapor interface. (b) Smooth surface tension function about the position. (c) Regions of the textured surface. (d) Constraints on the droplet. 1. Droplet Volume, 2. Penetration, 3. Element

length.

The line tension based front-tracking method is used to model the contact and detachment phenomena of droplet on textured surface. The free energy of the droplet can be depicted as follow[2].

$$E = \gamma_{LV}(A_{LV} - A_{SL}\cos\theta_e) + \rho Vgh_{cm}$$

Here, $\gamma_{LV}, \theta_e$ refer the surface tension and Young's angle of the substrate, which is the contact angle when the surface is flat. $A_{LV}, A_{SL}$ refer the interfacial area between liquid and vapor, and substrate and liquid. $\rho, V, g, h_{cm}$ refer the density of the liquid, volume of the liquid, gravitational acceleration, and height of the center of mass. In the present study, the droplet contour is modelled by a chain of nodes connected with line segments. The nodes represent the substrate refer the interface between liquid and substrate, and the others represent the interface between liquid and the vapor (**Fig. S2a**). Thereafter, we obtain the droplet shape by energy minimization procedure updating the node positions subjected to several constraints.

To enable gradient based optimization, we adopted smooth surface-liquid interaction function to guarantee the gradient of energy is differentiable for all possible configurations (**Fig. S2b**). The hyper tangent function is chosen as the surface-liquid interaction function as,

$$\gamma(y) = \frac{1}{2}(\gamma_1 + \gamma_2) + \frac{1}{2}(\gamma_1 - \gamma_2)\tanh(\alpha(y - CL)),$$

where $\alpha$ is coefficient determining how the surface tension is changed sharply, $\gamma_1, \gamma_2$ each refer $\gamma_{LV}$, and $\gamma_2 = \gamma_{SL} - \gamma_{SV} = -\gamma_{LV}\cos\theta_e$, and CL refers the contact line thickness, which determines the type of nodes. When nodes are outside of the CL zone, the surface tension of line segments is given as $\gamma_{LV}$, which is the value for liquid-vapor interface, when inside, the surface tension becomes $\gamma_{SL} - \gamma_{SV}$, which the value for liquid-substrate interface. The value smoothly changes near the contact line due to our choice of the hyper tangent function. After the surface-node interaction function is constructed, the interfacial energy of each node ($E_S$) can be computed by integrating the surface tension along the element line as follow (**Fig. 5.2a**).

$$E_S = \int_L \gamma(y) ds$$

$$= \sqrt{(x_2 - x_1)^2 + (y_2 - y_1)^2} \left( \frac{\gamma_1 + \gamma_2}{2} \right.$$

$$\left. + \frac{\gamma_1 - \gamma_2}{2} \frac{1}{\alpha(y_2 - y_1)} \ln\left(\frac{\cosh(\alpha(y_2 - CL))}{\cosh(\alpha(y_1 - CL))}\right) \right)$$

Here, $x_1, y_1, x_2, y_2$ refer the coordinate of each end of the element.

After the surface tension function for one surface is determined, we divided the textured geometry into several regions, which the surface tension function can be formulated by the function of *x, y* or *x+y*, and the surface energy of the droplet was computed by summing all the interfacial energy of elements (**Fig. S2c**). When a node crosses the region, the surface energy for each region were separately calculated and summed. When a node crosses more than one periodicity of the roughness, the surface energy for each periodicity was calculated separately and then summed.

Three constraints $(h_1, h_2, h_3)$ were adopted to model the shape of the droplet (**Fig. S2d**). First, the volume of the droplet should be conserved, during the energy minimization. Second, the penetration factor, defined as the squared sum of penetrated length ($d_i$), which is the distance to the closest substrate boundary should be zero. Finally, to prevent the element to be too short or long, a segment ($L_i$) longer or shorter than the maximum/minimum length ($L_{max}, L_{min}$) was squared.

$$h_1 = V - V_0$$

$$h_2 = \sum_{i=1}^{N} d_i^2$$

$$h_3 = \sum_{i=1}^{N} (L_i - k)^2$$

$$k = L_{min}(L_i < L_{min}), L_{max}(L_i > L_{max}), 0(L_{min} \leq L_i \leq L_{max})$$

The augmented Lagrangian method was used to minimize the free energy of the droplet while conserving the constraints. The total energy of the droplet including the Lagrangian term can be shown as, $E_{tot} = E_S + E_G + E_L$, where each term refer the surface energy, gravitational energy, and the Lagrangian energy. The Lagrangian energy can be depicted as,

$$E_L = \frac{1}{2} \sum_{j=1}^{3} (\mu_k^j h_j^2 + \lambda_k^j h_j)$$

Where, j,k refer to the kind of constraint and step number of minimization, and $\mu, \lambda$ refer to the Lagrangian coefficients. The minimization on the total energy of the droplet ($E_{tot}$) was conducted with conjugate gradient method, and the iteration was ended when the gradient of the total energy becomes sufficiently small. For every step of minimization, the Lagrangian coefficients were updated upon the minimization research. If the constraint factors were decreased enough, $\lambda_k^j$ was updated as,

$$\lambda_k^j = \lambda_{k+1}^j - r_k^j h_j,$$

and if the constraint factors became not small enough, the $r_k^j$ was updated as

$$r_{k+1}^j = 2 \times r_k^j$$

The minimization procedure was repeated until the convergence criteria based on gradient and the constraint are satisfied to reach sufficiently small gradient of total energy and constraint factors.

Between the energy minimization procedures, there were post process to accelerate the minimization procedure. For each minimization step, we test if the length of a line segment is longer than the maximum critical length or shorter than the minimum critical length. If a pair of nodes were too close to each other, the nodes between them were deleted to make the line segment length longer. If distance between a pair of nodes are too long, a new node were made between them to make the line segment length shorter.

Another procedure was employed to further enhance the accuracy of droplet shape prediction. If the length of a line segment is larger than the width of a pillar on the textured substrate, the nodes could not model the boundary between liquid and the textured substrate precisely. We could avoid this problem by restricting the maximum length of a line segment, however it made the computational cost to be significantly larger. Therefore, we used denser nodes near the substrate, and sparser nodes far from the substrate.

The numerical modeling enable us to seamlessly model the attachment and detachment of a droplet on a textured surface. We believe the numerical method can be used for many wetting phenomena on rough substrate such as contact angle hysteresis, sliding, wetting transition and droplet transportation. In the present study, we demonstrate one application of the numerical method to find the advancing and receding angle of the rough substrate in the presence of the gravity.

**Supplementary Note 3: Apparent contact angle at advancing and receding condition**

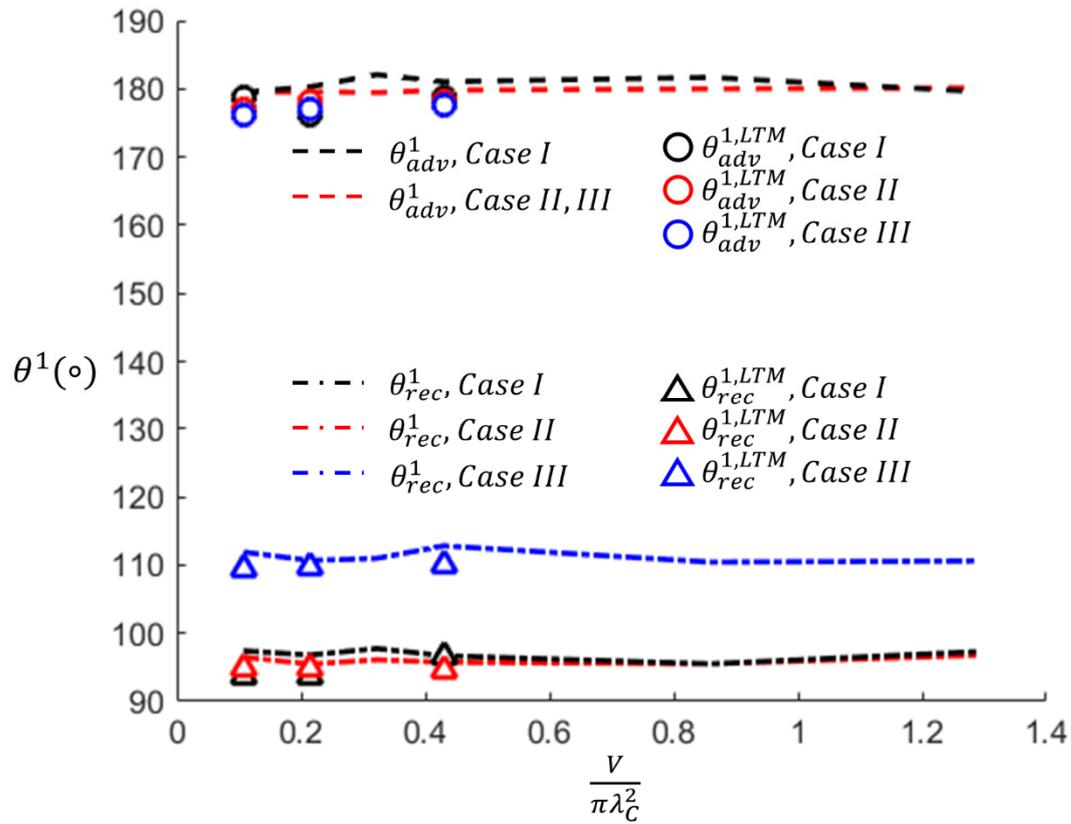

**Figure S3. Apparent advancing and receding angle of the droplet.** The apparent contact angle at advancing and receding condition is not affected by the gravity or volume of the droplet. Case I: ($\theta_e = 95°, W = 25\mu m, G = 100\mu m$), Case II: ($\theta_e = 95°, W = 25\mu m, G = 25\mu m$), Case III: ($\theta_e = 110°, W = 25\mu m, G = 25\mu m$)

While the zero-g angles at the advancing and receding conditions are affected by the gravity or the volume of the droplet, the apparent contact angle does not change both in advancing and receding condition, as shown in **Fig. S3**. The apparent advancing angle remains to be 180°, because the advancing angle turns out to be determined by geometrical interference criterion. The droplet which has apparent contact angle larger than 180° makes interference with pillars, so that the droplet cannot be stabilized. On the other hand, the apparent receding angle remains Young's angle, because the apparent contact angle of the droplet on flat surface is known to be not affected by the gravity[3], and the contact line starts to recedes when the apparent contact angle becomes smaller than the contact angle when the surface is flat.